\documentstyle[manuscript,aps]{revtex}
\begin{document}
\draft
\title{$I$-$V$ Characteristics of High Temperature Superconductors
with Columnar Defects}
\author{Mats Wallin and S. M. Girvin}
\address{Department of Physics, Indiana University, Bloomington IN 47405}
\date{\today}
\maketitle
\begin{abstract}
The vortex glass transition in the presence of columnar defects is
studied by Monte Carlo simulations of a vortex loop model,
suggested by the analogy to the $T=0$ superconductor-insulator
transition for dirty bosons in (2+1)D.
{}From finite-size scaling analysis of the $I$-$V$ characteristic we
find two dynamical exponents describing the non-equilibrium behavior.
We obtain $z_{\perp} = 6 \pm 0.5$ and $z_{\parallel} = 4 \pm 0.5$
when the current is applied perpendicular and parallel to the columnar
defects respectively.
\end{abstract}

Strong thermal fluctuations and quenched disorder give properties
to high temperature superconductors not observed in
conventional type II superconductors.
The irreversibility line separates two regions in the mixed state with
distinct properties of the penetrating magnetic field lines.
Above the irreversibility line, vortex lines are in a liquid state due to
the strong thermal fluctuations.
Matthew Fisher \cite{fisher} has suggested that in the presence of
quenched disorder there is a phase transition into a new superconducting
state at the irreversibility line, called the vortex glass \cite{shih}.
One signature of a vortex glass transition is universal scaling
of the nonlinear current-voltage ($I$-$V$) characteristic \cite{fisher}.
This has been verified in several
experiments \cite{koch,gammel,olsson,myoren,ando},
and in Monte Carlo simulations \cite{huse,reger,stroud,li}.

The original vortex glass model assumes pointlike impurities.
In a recent experiment Civale {\it et al.} \cite{civale} found a remarkable
shift upwards in the irreversibility line
of several Tesla upon irradiating a sample of
YBa$_2$Cu$_3$O$_7$ by heavy ions.
The damage tracks from the ions form permanent ``columnar defects''
with linear dimension comparable to the sample size.
These columnar defects strongly enhance flux pinning in the superconductor.

Nelson and Vinokur \cite{nelson-vinokur} derived scaling relations
for the $I$-$V$ characteristic for the case of columnar defects,
using an analogy to the Bose glass \cite{nelson,fisher-lee}.
In this picture, the glass transition is analogous to the
zero temperature superconductor-insulator
transition of dirty bosons in two dimensions.
The magnetic flux lines through the 3D bulk superconductor correspond
to world lines of bosons in the (2+1)D path integral formula for the
partition function. Static disorder for the (2+1)D boson system
corresponds to columnar defects for the 3D superconductor.
The superconducting phase corresponds to the
insulating localized Bose glass phase for dirty bosons,
and the resistive vortex fluid phase corresponds to the
superconducting phase of the dirty bosons.

A question of principal interest is whether a stable glass
phase can actually exist in 3D.
This issue was studied by Monte Carlo (MC) simulations of gauge glass
models with point disorder \cite{huse,reger,gingras}.
These found that the lower critical dimension is somewhat smaller than $d=3$,
but so close to $d=3$ that it is hard to resolve unambiguously.
On the other hand, the existence of the glass phase in the presence
of columnar defects is clear due to the analogy with the dirty boson
problem which has a phase transition which is now well
established \cite{fwgf,paper1}.
Furthermore, in these calculations screening of the
flux lines, and the anisotropy due to the preferred direction
picked out by the applied magnetic field, were not considered.

In this Letter we report Monte Carlo (MC) results for the nonlinear
$I$-$V$ characteristic at the glass transition of a model 3D
high temperature superconductor with columnar defects.
Screening of the flux lines is taken into account by neglecting
long-range vortex interactions altogether and assuming only a
short-range repulsion. Our model effectively assumes that the
magnetic field fluctuations diffuse more rapidly than the vortices.
The more difficult case of long-range forces and columnar defects
is being considered elsewhere \cite{lee-stroud-girvin}.

Let us briefly discuss scaling of the nonlinear $I$-$V$ characteristics
at the glass transition \cite{fisher,nelson-vinokur}.
Any configuration of wiggling vortex lines may be described as
a constant background of straight lines plus closed loops.
The characteristic size of the loops defines the correlation
length $\xi$, which diverges as $\xi \sim |T-T_g|^{-\nu}$.
The correlation time is assumed to obey standard dynamical scaling:
$\tau \sim \xi^z$,
where $z$ is the dynamical exponent which will be computed below.
Columnar defects make the system strongly anisotropic,
and this is reflected in the divergence of the correlation length.
{}From the analogy to the dirty boson problem we infer that the
correlation length in the direction of the columns scale as
$\xi_{\parallel} \sim \xi_{\perp}^{z_Q}$,
where $z_Q=2$ is the quantum ``dynamical
exponent'' of the dirty bosons at $T=0$ in 2D \cite{paper1}.

Consider voltage fluctuations across a correlation volume of size
$\xi_{\perp} \times \xi_{\perp} \times \xi_{\parallel}
\sim \xi_{\perp}^{2+z_Q}$.
The linear response electric field $E$,
the current density $I$, and the resistivity $\rho$ are related by
$E = \rho I$. From the Josephson and Nyquist relations the resistance
$R$ across the correlation volume scales as $R \sim \xi^{-z}$.
In the case of a transverse current applied in the direction
perpendicular to the columnar defects, the linear resistivity becomes
$\rho = R \xi_{\perp} \xi_{\parallel} / \xi_{\perp}
\sim \xi_{\perp}^{z_Q-z}$,
and $E \sim \xi_{\perp}^{z_Q-z} I$.
We also consider the case of an applied longitudinal current.
Here we find
$\rho = R \xi_{\perp}^2 / \xi_{\parallel} \sim \xi_{\perp}^{2-z_Q-z}$,
and $E \sim \xi_{\perp}^{2-z_Q-z} I$.

We now have to find the nonlinear $I$-$V$ characteristics.
At finite current density $I$, vortex lines fall out of equilibrium
due to the Lorentz force. The work against the Lorentz
force to create a closed loop costs energy $E \sim \pm AI$,
where $\pm$ is the orientation and $A$ the area of the loop.
For large enough loops the energy shift reaches $k_B T$.
This defines a characteristic ``current length'' $\xi_I$ beyond which
correlations are destroyed \cite{fisher}. In an isotropic system we
would define $\xi_I^2 I =1$ (dropping a factor of $k_B T$).
With columnar defects and a transverse current we instead take
$(\xi_{\perp} \xi_{\parallel})_I I \sim \xi_I^{1+z_Q} I = 1$.
The nonlinear $I$-$V$ characteristic is now expected to have the form
$V \sim I^{y_{\perp}}$ with exponent $y_{\perp}=(1+z)/(1+z_Q)$, and,
in general, the electric field is expected to scale like
$E \sim \xi_{\perp}^{-(1+z)} F_{\pm} (I\xi_{\perp}^{1+z_Q})$,
where $F_+$ and $F_-$
are universal scaling functions valid for $T \ge T_g$
and $T < T_g$, respectively \cite{nelson-vinokur}.
In the case of a longitudinal current we instead have
$(\xi_{\perp}^2)_I I \sim \xi_I^2 I = 1$, and following the same
derivation gives $V \sim I^{y_{\parallel}}$ with $y_{\parallel}=(z_Q+z)/2$,
and $E \sim \xi_{\perp}^{-(z_Q+z)} F_{\pm} (I\xi_{\perp}^2)$.

For a finite system at $T=T_g$ the diverging correlation length is cut
off by the system size $L$. With transverse $I$ we therefore expect
$E \sim L^{-(1+z)} F (IL^{1+z_Q}).$ Thus a plot of $E L^{1+z}$
versus $I L^{1+z_Q}$ is expected to give a universal, system size
independent curve. Notice that the meaning of the scaled quantities is
very natural. We have $E L^{1+z} \sim V / \langle V^2 \rangle^{1/2}_{eq}$,
where $V=EL$, $\langle V^2 \rangle^{1/2}_{eq}$ is the rms equilibrium voltage
fluctuation, and $I L^{1+z_Q} \sim (L/\xi_I)^{1+z_Q}$.

For our simulation we want a minimal model that describes the
collective behavior of interacting vortex lines in the presence of
thermal fluctuations and columnar defects.
We choose the dirty boson action \cite{paper1}
\begin{equation}
\beta H = \beta \sum_{\bf r} \Bigl\{ \frac{1}{2} {\bf J}^2({\bf r})
- v({\bf r}_{\perp}) J_z({\bf r}) \Bigr\} \; ,
\label{eq:H}
\end{equation}
where $\beta \equiv 1/k_B T$,
${\bf J} = (J_x,J_y,J_z)$ are integer variables
defined on the links of the lattice,
taking values from $-\infty$ to $\infty$.
These integer variables represent the vortex lines,
for example, $J_x({\bf r}) = \pm 1$ means that one vortex
line with orientation
$\pm 1$ goes from ${\bf r}$ to ${\bf r} + \hat{\bf x}$.
The vortex ``current'' is constrained to be divergenceless.
The first term in $H$ acts as a string tension and as
a short-range repulsion.
The second term in $H$ describes the columnar defects.
$v({\bf r}_{\perp})$ is a random site energy with uniform probability
distribution on the interval $[0,1]$.

Due to the anisotropy from the columnar defects,
the correlation length in the direction of the columns diverges at
$T_g$ as $\xi_{\parallel} \sim \xi_{\perp}^{z_Q}$ with $z_Q=2$.
To enable finite-size scaling we must use lattice sizes
that scale correspondingly.
We use lattices of size $L \times L \times L^2/4$
with periodic boundary conditions in all directions.
We use an applied magnetic field such that there is a fixed density
of $f=1/2$ of vortex lines in the direction of the columnar defects,
corresponding to half filling in the dirty boson problem.
The transition point $T_g = 0.248 \pm 0.002$ for this model is
already known from our previous work on the superconductor-insulator
transition \cite{paper1}.

Our MC algorithm consists of attempts to insert closed loops of vortex
lines with random sign on the plaquettes of the lattice.
The algorithm is ergodic in the sense that all possible configurations
of closed loops are accessible.
Non-equilibrium effects due to the Lorentz force from the finite current
are included by biasing the acceptance of plaquette moves. The energy
change of a move includes a term of the form
$\Delta \epsilon = ( \bbox{\nabla} \times
{\bf J} ) \cdot {\bf I}$,
where the discrete curl gives the orientation of the plaquette loop.

We do a thermal average over vortex loops for each realization of the
columnar disorder, followed by a disorder average.
As an estimate of the required number of MC update sweeps in the
thermal averages we use the time to return to steady state after
a reversal of the current.
The voltage is measured over up to $10^7$ sweeps,
with up to $10^6$ initial sweeps discarded to reach steady state.
For large currents a smaller number of sweeps is sufficient.
Also the number of disorder realizations necessary to get small
statistical error depends on the size of the applied current.
At large enough current, so that $\xi_I \ll L$, the system will self-average
and a small number of disorder realizations is enough.
At small $I$ we use up to 20 realizations of the disorder.

The voltage is given by the rate of phase slip across the system.
This is measured by keeping track of the number of plaquette
moves in the plane perpendicular to the applied current.
The electric field is proportional to
\begin{equation}
E \propto \frac{1}{\Omega} \left[ \left\langle \frac{d}{dt} ( N_+ - N_- )
\right\rangle \right] \; ,
\label{eq:E}
\end{equation}
where $\Omega=L_xL_yL_z$ is the volume of the system,
$N_{\pm}$ is the number of accepted plaquette moves with positive/negative
orientation in the plane perpendicular to the applied current,
and the time derivative stands for the change per sweep.
The bracket $\langle \; \rangle$ denotes thermal average,
and $[ \; ]$ denotes quenched disorder average.
Equation (\ref{eq:E}) implicitly assumes heavily overdamped
dynamics so that MC time can be equated with real time.
The update order in a sweep through the lattice is random rather than
sequential, in order to avoid possible systematic error at high current
where the acceptance rates are high.

Monte Carlo results for the nonlinear $I$-$V$ characteristic
at the glass transition are shown in Fig.\ \ref{fig:IV}.
In (a) the current density $I$ is transverse to the
columns, and in (b) $I$ is longitudinal \cite{footnote1}.
The dynamical exponent $z$ was adjusted until all data for different
lattice sizes collapse on the same curve.
Surprisingly, this happens for {\it different} values of $z$
for a transverse and a longitudinal current.
In (a) $z$ has been set to $z_{\perp}=6$, and
in (b) $z$ has been set to $z_{\parallel}=4$.
The MC data has three different regimes:
(1) For small current, where $\xi_I > L$,
the MC data points approach the dashed line, whose slope is 1.
This is a finite size effect: when the current length is cut off
by the system size $L$ the response becomes ohmic.
(2) In the nonlinear $I$-$V$ scaling regime, $1 < \xi_I < L$,
the MC data nearly coincides with the solid line, whose slope is given
by the scaling prediction for the exponent $y$ in $V \sim I^y$.
In (a) $y_{\perp}=(1+z_{\perp})/(1+z_Q)=7/3$,
and in (b) $y_{\parallel}=(z_Q+z_{\parallel})/2=3$.
(3) At higher current, with $\xi_I < 1$, scaling breaks down
as expected (data not shown).
Error bars on the MC points represent the standard deviation of the
voltage fluctuations between different disorder realizations.

What is the significance of the scaling plots in Fig.\ \ref{fig:IV}?
The critical temperature $T_g$ was determined
independently \cite{paper1}, and is not an adjustable parameter.
The figure shows that by adjusting only a {\it single} parameter,
the dynamical exponent $z$, we simultaneously get {\it two} things:
(1) all MC data collapse on the same universal curve, and
(2) the curve is a remarkably good power law over six decades with an
exponent which is consistent with the predictions of the scaling analysis.
For other choices of $z$ both these properties quickly break down.
{}From the rate at which the data collapse breaks down,
we estimate $z_{\perp}=6 \pm 0.5$ and $z_{\parallel}=4 \pm 0.5$.
One would have expected, at least in equilibrium,
that a single dynamical exponent would suffice,
but our simulation shows that in this non-equilibrium case we need two
\cite{footnote2}.
Attempts to use a single exponent $z \approx 5$ give a distinctly poorer
fit to the data.
As a test of universality we changed the aspect ratio of the
lattices from $L_z = L_x^2/4$ to both $L_x^2/8$ and $L_x^2$.
We also varied the applied magnetic field from $f=1/2$ to $f=1/4$.
These did not change the values of the dynamical exponents.
We have not yet tested lattice structures other than simple cubic.

The following simple argument suggests how the two different
dynamical exponents $z_{\perp} = 6$ and $z_{\parallel} = 4$ might arise.
First consider an isotropic system,
and study the motion of a blob of vortex lines of diameter $\xi$.
As a crude approximation we view the blob as a polymer containing
$N \sim \xi^2$ segments. The time it takes each segment to random walk a
distance $\xi$ scales as $t_1 \sim \xi^2$. The diffusion coefficient for
the center of mass (CM) is down by a factor of $N$. Hence the time for the CM
to move a distance $\xi$ scales as $t \sim \xi^4$, and thus $z=4$.
This argument neglects correlations in the motion of the segments of
the vortex lines.
The result coincides with the mean field result \cite{dorsey}.
When a finite current is applied, the role of the diverging correlation
length $\xi$ is taken over by the finite current length $\xi_I$,
which in the isotropic case is defined by
$\xi_I^{1+1} I = 1$. The time for the blob to move is
now $t \sim \xi_I^{2(1+1)}$, so again $z=2(1+1)=4$.
In the anisotropic case with columnar defects and a transverse current,
the current length is defined by $\xi_I^{1+z_Q} I = 1$.
This is now the characteristic length that the blob has to move.
If we assume that the time for the CM of the blob to move is now
$t \sim \xi_I^{2(1+z_Q)}$, we find $z_{\perp}=2(1+z_Q)=6$.
With a longitudinal current
we instead have $\xi_I^{1+1} I = 1$,
and the correlation time is $t \sim \xi_I^{2(1+1)}$,
so $z_{\parallel}=2(1+1)=4$.
If our assumption is justified,
this argument could explain why the non-equilibrium correlation time
is different when the current is applied in different directions.
Exactly what happens when the applied current goes to zero is
not clear from this argument,
however one normally would expect a single relaxation time.
We note in passing that a measurement of the voltage noise spectrum
at the critical point
should be able to independently determine the predicted current
dependence of the characteristic frequency scales in the system.

How do our results compare with experiments on systems without
deliberately introduced columnar defects?
Experiments \cite{bishop} on single crystals of YBCO give
$z=4.5 \pm 1.5$ and $\nu = 2.0 \pm 1$,
by fitting data to point-disorder vortex glass scaling laws.
In these experiments correlated disorder might be present in
the form of dislocations, twin boundaries, etc.
If, following Nelson and Vinokur \cite{nelson-vinokur}, we instead
fit to the Bose glass scaling laws used in this paper for
columnar defects, we find $z=7 \pm 2$ and $\nu = 1.3 \pm 0.5$.
These are roughly consistent with our values $z_{\perp} = 6 \pm 0.5$
and $\nu = 1.0 \pm 0.1$, where $\nu$ was calculated in previous
equilibrium simulations of dirty bosons \cite{paper1}.
Simulations of point and twin-boundary disorder is underway.

In summary, we report Monte Carlo simulations of the nonlinear
$I$-$V$ characteristic of a model high-temperature superconductor
with columnar defects.
We find two dynamical exponents describing the non-equilibrium behavior,
$z_{\perp} = 6 \pm 0.5$ and $z_{\parallel} = 4 \pm 0.5$ in the case
of an applied transverse and longitudinal current respectively.
Experimental measurements in systems with deliberately introduced columnar
defects would be highly desirable.

Many helpful discussions with
D. Nelson, M. P. A. Fisher, D. Stroud, D. Huse, A. Sudb\o,
A. P. Young, N. Goldenfeld, and A. Dorsey are gratefully acknowledged.
MW is supported by NFR grant F-DP 9902-303.
SMG is supported by DOE grant DE-FG02-90ER45427 and NCSA DMR-910014N.

\newpage

\begin{figure}
\caption{Log-log plot of Monte Carlo results for the nonlinear $I$-$V$
characteristic at the vortex glass transition with columnar defects.
Plotted in (a) is MC data for $EL^{1+z}$ versus $IL^{1+z_Q}$,
where $E$ is the electric field, and the applied current density $I$
is perpendicular to the columnar defects.
Plotted in (b) is $EL^{z_Q+z}$ versus $IL^2$, where the current
density $I$ is parallel to the columns.
The dynamical exponent $z$ has been adjusted until all data
for different system sizes collapse on the same curve.
In (a) $z$ has been set to $z_{\perp} = 6$, and in (b) $z_{\parallel} = 4$.
For other values of $z$ scaling quickly breaks down.
The solid straight line in (a) has slope
$y_{\perp} = (1+z_{\perp})/(1+z_Q) = 7/3$ given by scaling (see text).
The solid straight line in (b) has slope
$y_{\parallel} = (z_Q+z_{\parallel})/2 = 3$.
The dashed straight lines in (a) and (b) both have slope $1$,
and indicate the ohmic region at small current.
\label{fig:IV}}
\end{figure}


\begin{references}
\bibitem{footnote1}
In order to improve the acceptance rate
the calculation in the case $I=I_{\parallel}$ was performed with one
flux line penetrating in the direction perpendicular to the columns.
This appears not to affect the slope of the curve.
\end{references}
\end{document}